\documentclass[aip,rsi,amsmath,amssymb,reprint]{revtex4-1}

\usepackage{graphicx,color}
\usepackage{amsmath,amssymb}
\usepackage{verbatim}
\usepackage{ulem}
\usepackage{mathptmx}

\long\def\symbolfootnote[#1]#2{\begingroup%
\def\thefootnote{\fnsymbol{footnote}}\footnote[#1]{#2}\endgroup} 
\def\blfootnote{\xdef\@thefnmark{}\@footnotetext}

\definecolor{purp}{rgb}{0.5,0,0.5}
\definecolor{darkgreen}{rgb}{0.1,0.7,0}
\definecolor{orange}{rgb}{1,0.5,0.2}
\definecolor{violet}{rgb}{0.7,0,0.7}
\definecolor{Cyan}{rgb}{0,0.5,0.5}

\newcommand{\be}{\begin{equation}}
\newcommand{\ee}{\end{equation}}
\newcommand{\bi}{\begin{itemize}}
\newcommand{\ei}{\end{itemize}}
\newcommand{\bea}{\begin{eqnarray}}
\newcommand{\eea}{\end{eqnarray}}

\begin{document}

\title{Theory of relativistic radiation reflection from plasmas} 
\author{Arkady~Gonoskov}
\affiliation{Department of Physics, Chalmers University of Technology, SE-41296 Gothenburg, Sweden}
\affiliation{Institute of Applied Physics, RAS, Nizhny Novgorod 603950, Russia}
\affiliation{Lobachevsky State University of Nizhni Novgorod, Nizhny Novgorod 603950, Russia}

\begin{abstract}
We consider the reflection of relativistically strong radiation from plasma and identify the physical origin of the electrons' tendency to form a thin sheet, which maintains its localisation throughout its motion. Thereby we justify the principle of the \textit{relativistic electronic spring} (RES) proposed in [A.~Gonoskov et al. PRE \textbf{84}, 046403 (2011)]. Using the RES principle we derive a closed set of differential equations that describe the reflection of radiation with arbitrary variation of polarization and intensity from plasma with arbitrary density profile for arbitrary angle of incidence. PIC simulations show that the theory captures the essence of the plasma dynamics. In particular, it can be applied for the studies of plasma heating and surface high-harmonic generation with intense lasers.

\end{abstract}

\maketitle

\section{Introduction}

The reflection of electromagnetic radiation from a plasma with overcritical density originates from the induced self-consistent dynamics of electrons at the plasma interface. If the radiation is intense enough to make the electrons' motion relativistic, the radiation pressure causes an inward relocation of electrons and enables a large variety of highly nonlinear reflection scenarios. These span between the cases of ideal reflection (the limit of highly overdense plasma with steep distribution) and relativistic self-induced transparency. Such Relativistic intensities can be achieved with high-intensity optical laser pulses, while overdense plasma with various scales of density transition at the interface is naturally formed by the ionization, heating and thermal expansion of solids exposed to pre-pulse light. The prospects of using laser-solid interactions for various applications, ranging from high-harmonic generation to plasma heating, has stimulated theoretical and experimental studies of the non-linear reflection process \cite{quere.prl.2006, dromey.nphys.2006, thaury.nphys.2007, dromey.nphys.2009, teubner.rmp.2009, heissler.apb.2010, heissler.prl.2012, mikhailova.prl.2012, rdel.prl.2012, vincenti.ncom.2014, riconda.ppcf.2014, fiore.pop.2014, ma.oe.2016, edwards.pra.2016, edwards.prl.2016, strelkov.ufn.2016, leblanc.nphys.2017, chatziathanasiou.photonics.2017}.

The most general theoretical description of the reflection process is given by the kinetic approach. Although this description is very useful for numerical studies, the high degree of nonlinearity largely precludes direct theoretical analysis based on the kinetic equations. A notable exception is the case of normal incidence of circularly polarized radiation. In this case, the balance between the radiation pressure and Coulomb attraction to the ions leads to quasi-stationary plasma distributions. These distributions can be obtained analytically in the hydrodynamical approximation \cite{kim.jetpl.2000, cattani.pre.2000}. However, in other cases the radiation pressure oscillates in time and gives rise to complex plasma dynamics. Some theoretical analysis can be performed in the limit of high density using the cold fluid approximation \cite{sanz.pre.2012, debayle.pop.2013, debayle.pre.2015}. However, in the general case oscillation of the radiation pressure leads to the formation of many streams in plasmas invalidating the hydrodynamical approximation. 

An alternative approach is to develop a simple artificial system, the behaviour of which mimics plasma dynamics in certain respects. The description in this case can be driven by phenomenological, rather than \textit{ab initio}, principles. If the plasma has a sharp boundary with steep rise of density to a sufficiently high value, the incident radiation penetrates to a negligible depth, and the deviation from ideal reflection can be modelled using the principle of the \textit{relativistic oscillating mirror} (ROM) \cite{bulanov.pop.1994, lichters.pop.1996, gordienko.prl.2004}. This principle is that the ideal reflection happens at some oscillating point, where the incoming and outgoing electromagnetic fluxes are equal to each other (Leontovich boundary conditions). Theoretical analysis based on the ROM principle provides insights into various aspects of interactions, such as polarization selection rules \cite{lichters.pop.1996, vonderLinde.apb.1996, teubner.rmp.2009} and high-harmonic generation properties \cite{baeva.pre.2006, pirozhkov.pop.2006}.
 
However, the assumed-to-be instantaneous redirection of the incident electromagnetic flux implies that the energy is not accumulated even for a fraction of the radiation cycle when the electrons are relocated by the radiation relative to the ions. Thus the ROM model cannot encompass effects due to significant electron displacement, which happens when the intensity is not too low and/or the density is not too high. Indeed, the boundary conditions in the ROM model explicitly imply that the amplitude of the reflected radiation can never exceed that of the incident radiation. However, for certain parameters, the electron displacement leads to the accumulation of up to 60\% of the energy of each radiation cycle, followed by the release of that energy in the form of an short burst with more than a hundred times higher intensity \cite{gonoskov.pre.2011}. A principle that accounts for such energy redistribution and describes this and other highly nonlinear interaction scenarios was proposed in Ref. \cite{gonoskov.pre.2011} and is known as the \textit{relativistic electronic spring} (RES). The RES model provides a direct description of the plasma and electromagnetic field dynamics over a large range of intensities and densities, which is thus referred to as the RES regime. Recent studies have showed that the RES regime is efficient at converting the energy of the incident radiation into coherent XUV bursts with short duration, high intensity \cite{bashinov.epjst.2014, fuchs.epjst.2014, blackburn.arxiv.2017} and controllable ellipticity \cite{blanco.arxiv.2017}, as well as for producing incoherent X-ray and gamma radiation \cite{nerush.pop.2014, serebryakov.pop.2015}.

In this paper, we reveal the physical origins of the RES principle and elaborate further the theory based on this principle. We provide general equations that are applicable for arbitrary incidence angle, arbitrary density profile and arbitrary temporal evolution of the field and polarization in the incident radiation. In this way, we demonstrate that the RES model does not just mimic the reflection process, but is a theory that arises from a physically-grounded approximation.

\section{Origins of the governing principles}

The primary assumption of the theory is that the plasma eventually halts the propagation of the incident radiation. This generally happens when the frequency range of the incident radiation is below the plasma frequency. If sufficiently high densities are reached at some point inside the plasma, then the radiation propagation is generally halted. Although effects of relativistic self-induced transparency requires more detailed analysis, here we assume that the density grows at the interface to sufficiently high values to prevent the radiation propagation. Under this assumption, we focus on the origins of the RES principle and answer the following questions: Why do the electrons tend to form a thin sheet? Why do the electrons maintain and sometimes even improve their co-locality in space during the motion of the sheet? Does the RES principle provide a self-consistent description of plasma dynamics under certain assumptions?

We consider the problem in the reference frame moving with velocity $c\sin\theta$ along the plasma surface, where $c$ is the speed of light and $\theta$ is the angle of incidence. In this reference frame the incidence is normal and the plasma streams with transverse speed of $c\sin\theta$. Under the assumption that the spatial scales of transverse variations of radiation and plasma are large in comparison with the wavelength, the problem can be locally considered as one-dimensional. When the incident radiation reaches the plasma, electrons start to move under the effect of the electromagnetic fields, while the same fields are modified by the induced electron and ion currents as they propagate deeper and deeper. However, the fact that the propagation of radiation is eventually halted means that the inward emission due to these currents must, at some point, provide exact cancellation of the incident radiation. Thus, the incident field cancellation by the induced currents is a general formulation of the radiation reflection. This cancellation requirement is one of the assumptions of the RES theory. 

One might expect the electron spatial distribution, which is determined by the self-consistent electromagnetic fields, to be highly complex. However, a remarkable simplification takes place in the case of relativistic motion: the electrons tend to form a single thin sheet that separates the region of uncompensated ions and the unperturbed plasma.  

We observed this tendency and its connection with relativistic effects in the consideration of the stationary problem in Ref.~\cite{gonoskov.pre.2011}. However, this does not explain why it occurs in the general dynamical case: although the electrons can naturally pile up into a localized bulk at the rising edge of the radiation pressure, one could expect that the opposite process, i.e. spreading, happens, when the radiation pressure decreases and the bulk propagates backwards. However, as one can see from fig.~3~(a) in Ref. \cite{gonoskov.pre.2011}, the bulk actually shrinks even further during this process. This gives rise to the generation of short bursts of radiation. In terms of the acting forces and the consequent particle dynamics, this effect can be qualitatively explained as follows.

We divide the motion of the electrons during a single cycle of radiation pressure oscillations into two stages: first, the radiation pushes electrons from left to right; then in the second stage, the formed bulk propagates from right to left (towards the initial position of the plasma boundary). During the first stage, at each instance of time the following statement holds true: the electrons in the left part of the bulk experience a stronger force of radiation pressure for longer time than the electrons in the right part of the bulk. If the force causes relativistic motion of electrons, then these difference quickly results in piling up the electrons. 

During the second stage, the mechanism by which the sheet becomes thinner is different. To demonstrate the idea we assume that the density of electrons $n$ is constant across the bulk and that the electrons move with roughly the same speed in the transverse direction (the difference cannot be dramatic because their motion approaches the relativistic limit). We use $x_r$ to denote the distance between a certain point within the bulk and the rightmost side of the bulk. In this case, with increase of $x_r$ the transverse component of the magnetic field $B_{\bot}$ and the related component of the Lorentz force grow linearly:  
\begin{equation}
B_{\bot} \sim n x_r,
\end{equation}
where $e$ is the electron charge. The electric component of the Lorentz force also grows linearly with increase of $x_r$. When the attraction to the residual ions start to dominate over the radiation pressure, the imbalance also grows linearly with increase of $x_r$:  
\begin{equation}
F_x \sim n x_r.
\end{equation} 
From the conservation of the canonical momentum we can conclude that the transverse momentum of electrons grows quadratically with $x_r$ (here we assume $p_{\bot} \gg mc$):
\begin{equation}
p_{\bot} \approx \frac{e}{c} \int_0^x B_{\bot}(x^\prime) dx^\prime \sim n x_r^2.
\end{equation}
Thus the electrons in the left part of the bulk have larger values of transverse momentum and are therefore more `massive' in terms of longitudinal motion due to relativistic increase of the effective mass. In the highly relativistic case, the effective mass for longitudinal motion grows quadratically with $x_r$: 
\begin{equation}
m_{\parallel} = m\sqrt{1 + p_{\bot}^2/(mc)^2} \sim n x_r^2.
\end{equation}
As we can see, with increase of $x_r$ the relativistic increase of mass grows quadratically, whereas the longitudinal force grows linearly. This means that the response to the restoring force of the electrons in the left part of the bulk is retarded relative to those in the right part. As a result, the electrons in the right part move to the left faster than the electrons in the left part. However, the electrons from the right part can never overrun the electrons from the left part. This is because of the conservation of transverse canonical momentum. Suppose some electron $L$ had initial position to the left of some electron $R$ within the bulk and further that the electron $R$ overruns the electron $L$. Then at some instance of time the electrons have the same longitudinal coordinate. At this instance of time the electrons have exactly the same transverse momentum, because this depends only on the longitudinal coordinate due to the conservation of transverse canonical momentum. However, prior to this instance the electron $R$ experienced a strictly weaker longitudinal force and thus gained less longitudinal momentum than the electron $L$. Thus, the electron $R$ has strictly smaller longitudinal velocity than the electron $R$. This contradicts the initial supposition that the electron $R$ overruns the electron $L$. The consequence of this is that the electrons in the bulk can come closer to each other but the effect of wave breaking can never happen. 

In such a way we showed that in the case of relativistic motion, the relativistic mass increase due to transverse momentum causes inversion of longitudinal velocities in the bulk, while the conservation of transverse canonical momentum prevents breaking of this inversion. Therefore the bulk tends to shrink during its backward motion. This means that the electrons in the bulk tend towards having the same longitudinal velocity. Since their motion is relativistic, and the orientation of the transverse motion is roughly the same (being determined by the magnetic field orientation), the transverse components of the electrons' velocities are roughly the same for all electrons within the bulk. This provides complete self-consistency with the macroscopic assumptions of the RES theory. In the RES theory we make use of the fact that the emission is determined by the electrons velocity but not momentum. Thus, although the electrons in the bulk do have different values of momentum, their emission can be described in terms of macroscopic parameters: the bulk's charge and velocity. 

The only special point in this respect is the point when the bulk moves almost exactly to the left. In this case the backward emission becomes singularly strong. The actual limit depends on the gamma factor distribution and the thickness of the bulk. Determining the actual limit of the bulk's shrinking requires consideration of its microscopic dynamics. The driving conditions for these dynamics can be obtained from its macroscopic dynamics described by the RES theory under the assumption of the bulk being thin in comparison with its macroscopic motion.

\section{Governing equations}

Here we again use the moving reference frame, where the incidence is normal and the problem can be considered as one-dimensional. Although we use here this approximation, the developed approach can be extended to account for various deviations from one-dimensional geometry. We also assume here that ions remain immobile, but their motion can be accounted for, for example, as a slow deviation to the presented consideration. We use orthogonal coordinate system XYZ with the $x$-axis oriented towards the incidence direction and $y$-axis oriented against the plasma stream. 

According to the RES principle, at each instant of time the plasma is assumed to consist of three regions: (1) a region $x < x_s$ that contains only plasma ions but no electrons; (2) a thin sheet of electrons at $x = x_s$, where the uncompensated charge of the first region is concentrated; and (3) unperturbed plasma for $x > x_s$.  The RES principle states that the radiation of the electrons in the thin layer and of uncompensated ions provides compensation of the incident radiation:
\begin{eqnarray}\label{s0}
	&& {E_y\left(\cos\theta\left(x_s - ct\right)\right) = \frac{2 \pi e}{\cos^2\theta} \int\displaylimits_{-\infty}^{x_s} N(x)dx \left(\sin\theta - \frac{\beta_y}{1 - \beta_x}\right),}\\
	&& {E_z\left(\cos\theta\left(x_s - ct\right)\right) = \frac{2 \pi e}{\cos^2\theta} \int\displaylimits_{-\infty}^{x_s} N(x)dx \left( - \frac{\beta_z}{1 - \beta_x}\right)},
\end{eqnarray}
where the arbitrary incident radiation is characterized in the laboratory frame through the electric field (in CGS units) in the plane of incidence $E_y(\eta)$ and in the other transverse direction $E_z(\eta)$ as functions of coordinate $\eta$ along the propagation (for the respective components of the magnetic field this implies $B_z(\eta) = E_y(\eta)$ and $B_y(\eta) = -E_z(\eta)$); the plasma is characterized by an arbitrary function $N(\chi)$ of density in the laboratory frame as a function of depth $\chi$; $\beta_x$, $\beta_y$ and $\beta_z$ are the effective (averaged) components of the electrons velocity in the sheet given in the units of the speed of light. If the fields are sufficiently strong to cause relativistic motion of electrons, the limit $\beta_x^2 + \beta_y^2 + \beta_z^2 = 1$ can be used to account for relativistic restriction. Note that the relativistic gamma factor does not directly enter the expressions for the layer emission. In some cases it might be important to consider the finite value of the gamma factor, however, as we understood above, the gamma factor is different for different electrons in the bulk. Thus the above-mentioned relativistic limit provides a natural self-consistent description in a simple form. Using that $q_s = 2 \pi e \int_{-\infty}^{x_s} N(x) dx / \cos^2\theta$ characterizes the instantaneous total charge of electrons in the layer, we can now write a closed system of differential equations that describe the reflection process:
\begin{eqnarray}\label{s1}
	&& {E_y\left(\cos\theta\left(x_s - ct\right)\right) = q_s \left(\sin\theta - \frac{\beta_y}{1 - \beta_x}\right),}\\
	&& {E_z\left(\cos\theta\left(x_s - ct\right)\right) = q_s \left( - \frac{\beta_z}{1 - \beta_x}\right)}, \\
	&& {\beta_x^2 + \beta_y^2 + \beta_z^2 = 1},\\
	&& {\frac{dq_s}{dt} = \frac{2 \pi e c}{\cos^2\theta} N(x_s) \beta_x},\\
	&& {\frac{dx_s}{dt} = c \beta_x}.
\end{eqnarray}
During the reflection process the backward emission appears as the component of the radiation of the uncompensated ions and the electrons in the layer in the negative $x$ direction:
\begin{eqnarray}\label{br}
	&& {E_y^b\left(\cos\theta\left(x_s + ct\right)\right) = q_s \left(\frac{\beta_y}{1 + \beta_x} - \sin\theta \right),}\\
	&& {E_z^b\left(\cos\theta\left(x_s + ct\right)\right) = q_s \left(\frac{\beta_z}{1 + \beta_x}\right)},
\end{eqnarray}
where the backward radiation is characterized in the laboratory frame through the electric field in the plane of incidence $E^b_y(\xi)$ and in the other transverse direction $E^b_z(\xi)$ as functions of coordinate $\xi$ along the specular direction (for the respective components of the magnetic field this implies $B^b_z(\xi) = -E^b_y(\xi)$ and $B^b_y(\xi) = E_z(\xi)$).

We can now show that the system (7-11) always provides exactly one solution and that this solution is physically meaningful. From the first three equations we can explicitly obtain:
\begin{equation}\label{beta_x}
\beta_x = \frac{R_y^2 + R_z^2 - 1}{R_y^2 + R_z^2 + 1},
\end{equation}
where the quantities $R_y = \beta_y/(1 - \beta_x)$ and $R_z = \beta_z/(1 - \beta_x)$ are given by:
\begin{eqnarray}\label{R_yz}
&& {R_y = \sin\theta -\frac{E_y\left(\cos\theta\left(x_s - ct\right)\right)}{q_s},} \\
&& {R_z = -\frac{E_z\left(\cos\theta\left(x_s - ct\right)\right)}{q_s}.}
\end{eqnarray}
The expression (\ref{beta_x}) always provides a value within a meaningful range $-1 < \beta_x < 1$. Another requirement for the solution to be meaningful occurs under the assumption that the plasma has a certain bound, which we can assume to be at $x = 0$, i.e. $N(x < 0) = 0$, $N(x > 0) > 0$. In this case, the solution has a physical meaning only if $x_s > 0$. We can show that this is always the case. If the value of $x_s$ approaches the point $x = 0$, the value of $q_s$ also tends to zero. In this case, according to Eqs.~(11, 12), the value $R_y^2 + R_z^2$ tends to grow (if $E_y \neq 0$ or $E_z \neq 0$). This eventually leads to $\beta_x > 0$ (see Eq.~\ref{beta_x}), precluding reaching the point $x = 0$. The only exception is the case when both $E_y = 0$ and $E_z = 0$. This can happen when the polarization is strictly linear. In this case, one can consider a linear approximation of the field in the vicinity of the zero point and demonstrate that the linearised equations always give a positive solution for $\beta_x$. Thus passing this special point implies that $\beta_x$ switches from negative to positive instantly at $x = 0$. (This result is expected, since we can always introduce a small deviation from linear polarization to resolve this special point and then consider the limit of the deviation to be infinitely small.)   

We have demonstrated that the theory always provides exactly one solution and this solution is physically meaningful. Then Eqs. (14-16) provide a practical means of computing the solution numerically. Assuming that at the instance of time $t = 0$ the incident radiation reaches the plasma at the point $x = 0$ (i.e. $E_y(\eta > 0) = 0$, $E_z(\eta > 0) = 0$, $N(d < 0) = 0$), we can write the initial conditions for the system (\ref{s1}) in the form:
\begin{eqnarray}\label{ic}
&& {x_s(t = 0) = 0,} \\
&& {q_s(t = 0) = 0.}
\end{eqnarray}
The Eqs.~(14-16) coupled with Eqs.~(10, 11) explicitly determine how $x_s$ and $q_s$ evolve, provided that we start from any negligibly small but non-zero values of $x_s$ and $-q_s$ (which is justified by the interest in the solution with a physical meaning). For practical reasons one can also avoid the aforementioned singular point by introducing a small deviation to the field in the points where $E_y = 0$ and $E_z = 0$. These practically motivated deviations do not affect the results. 

\section{Comparison with numerical simulations}

To demonstrate the capabilities of the theory we present a comparison of the theoretical results with the results of a PIC simulation for a particular interaction scenario. We consider a pulse of radiation incident on a plasma slab with incidence angle of $\theta = \pi/7$. We again consider the problem in the moving reference frame, where the problem is one-dimensional. In this reference frame the pulse is defined by the field components $E_y(x - t) = B_z(x-t) = 300\sin\left(x-t\right)$, $E_z(x - t) = -B_y(x-t) = 150\sin\left(\left(x-t\right)7/4\right)$, where coordinate $x$ and time $t$ are given in the units of $\lambda/\left(2\pi\right)$ and $\lambda/\left(2\pi c\right)$ respectively; and the field strength is given in relativistic units $2 \pi c^2/\left(e \lambda\right)$, $\lambda = (1\, \mu \text{m}) /\cos\theta$. The plasma is comprised of immobile ions and electrons with the density rising linearly from 0 to 500 over $0 < x < \lambda/3$, staying fixed over $\lambda/3 < x < 2\lambda/3$ and falling linearly to 0 over $2\lambda/3 < x < \lambda$. Here the density is given in units of $n_{\text{cr}}/\cos\theta$, where $n_{\text{cr}} = \pi m c^2/\left(e \lambda\right)^2$ is the plasma critical density for the wavelength $\lambda$ in the laboratory frame.

\begin{figure}[htbp]
		\centering
			\includegraphics[width=\linewidth]{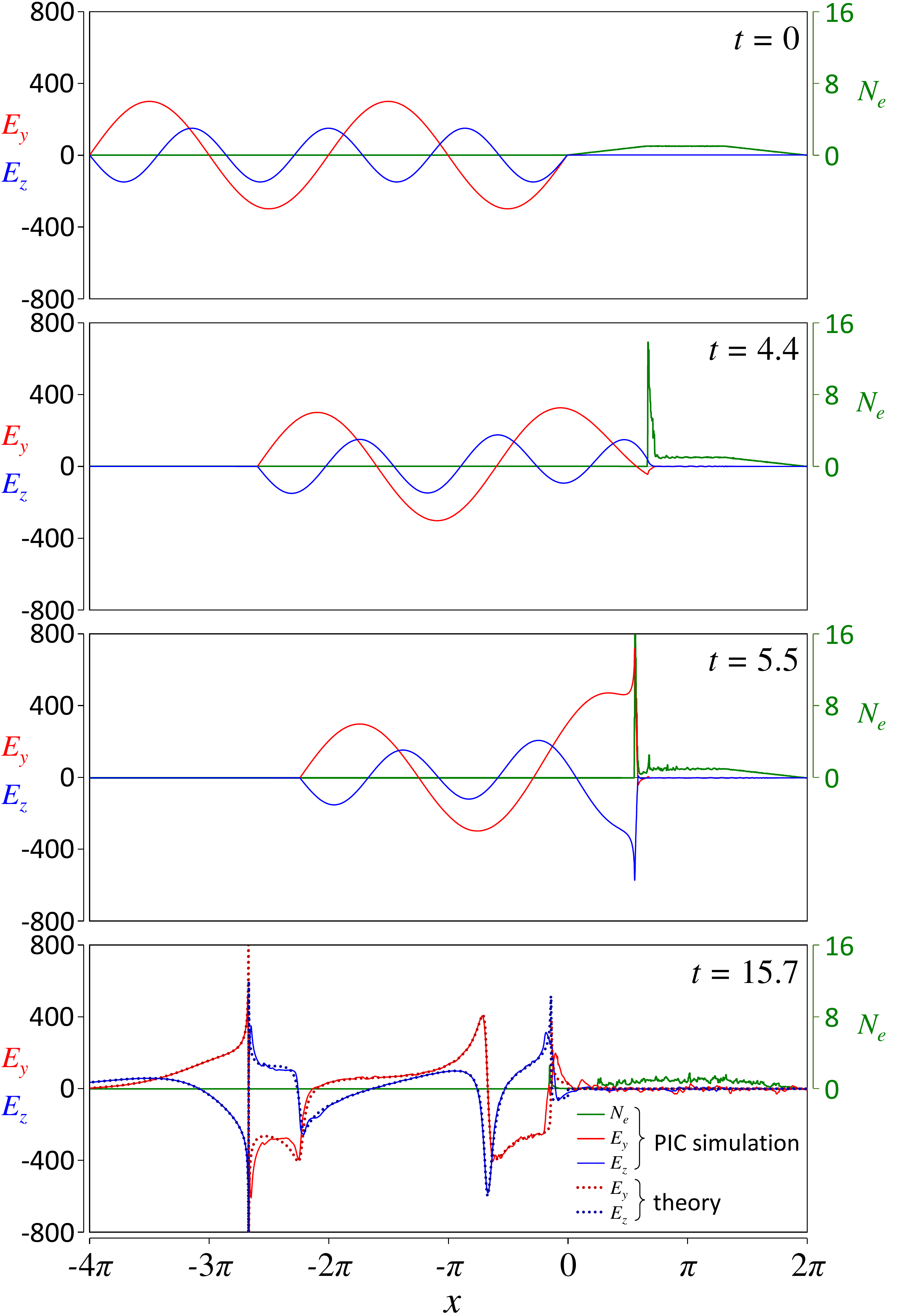}
			\caption{Comparison of theoretical calculations with the result of PIC simulation for the scenario described in the text. The panels show the electric field $y$- and $z$-component and the electron density as a function of longitudinal coordinate in the moving reference frame for four instances of time: $t = 0$ (initial distributions), $t = 4.4$ (the radiation pushing electrons that pile up into a bulk), $t = 5.5$ (the bulk shrinking during backward motion), $t = 15.7$ (the resulted reflected signal that propagates from right to left). The result of numerical integration of the theory equations are shown with dotted curves for $t = 15.7$.}
		\label{fig:fig1}
	\end{figure}

The results of 1D PIC simulation for this problem are shown in fig.~1 for four instances of time. At the instance $t = 4.4$ we can clearly see how the incident radiation pushes electrons so that they form a dense bulk. At the instance $t = 5.5$ we can see how this bulk shrinks further during its backward motion and how this results in the generation of a singular burst of radiation. At the last instance $t = 15.7$ we can see the resulting reflected radiation. The result of numerical integration of the equations (7-13) is shown in terms of $E_y(x+t) = -B_z(x+t)$ and $E_z(x+t) = B_y(x+t)$ with dotted curves. As we can see, the theory describes the entire process well. The most difficult instance for the theory is the instance of the burst generation, when $\beta_x \approx -1$. At this point, the theory gives singular results because the gamma factor is assumed to be infinite. The results are not so sensitive to this assumption at other instances of time. The analysis presented above shows that it is not reasonable to consider any particular value of gamma factor, because it is different for different electrons within the bulk. This point is of particular interest for the generation of short bursts of radiation and plasma heating because the electron bulk undergoes the most extreme bifurcation. To study these problems one needs to consider micro-dynamics of the electron bulk. The presented theory can be very useful for determining the macroscopic conditions for these studies. 

As one can see from the picture for $t = 15.7$ after the singular point at $x \approx -2.6\pi$ the resulted emission starts to deviate slightly (in a non-systematic but rather random way) from the predictions of the theory at $x > -2.6\pi$. However, these deviations quickly decay and the generated signal again follows perfectly the prediction from $x > -2\pi$. This indicates that the theory encompasses the essence of the plasma dynamics, while the particular perturbations decay quickly so that the plasma does not accumulate and "remember" earlier deviations. The parameters of the considered example have been chosen arbitrarily; similar or even better agreement can be seen in other cases. 

\section{Conclusions}

In this paper, we have identified the physical origins of the RES principles and demonstrated that these principles emerge from the general tendency of electrons to bunch into a thin sheet due to relativistic effects in radiation-plasma interactions of arbitrary type. Using the RES principles, we developed a theory that is capable of describing radiation-plasma interactions for arbitrary variation of polarization and intensity in the incident radiation, arbitrary density profile of the irradiated plasma, as well as arbitrary angle of incidence. The theory can be applied for studies of surface high-harmonic generation and plasma heating with intense lasers. It can also guide theoretical and experimental studies by revealing the dependence of interaction scenarios on the incidence angle, shape of the plasma density profile, as well as laser pulse shape, intensity, ellipticity, and carrier-envelope phase. 

\section{Acknowledgements}
The author would like to thank M.~Marklund and T.~G.~Blackburn for useful discussions. The research is supported by the Russian Science Foundation Project No. 16-12-10486, and by the Knut \& Alice Wallenberg Foundation.

\bibliography{literature}
 \bibliographystyle{aipnum4-1}

\end{document}